\documentclass[preprint]{revtex4}
\usepackage{bm,color}
\usepackage{graphicx}
\usepackage{amsmath}

\begin{document}
\title{Current-induced skyrmion dynamics in constricted geometries}
\author{Junichi Iwasaki$^{1}$}
\author{Masahito Mochizuki$^{2}$}
\author{Naoto Nagaosa$^{1,3}$}
\email{nagaosa@riken.jp}
\affiliation{$^1$ Department of Applied Physics, The University of Tokyo, 7-3-1, 
Hongo, Bunkyo-ku, Tokyo 113-8656, Japan}
\affiliation{$^2$ Department of Physics and Mathematics, Aoyama Gakuin University, 5-10-1, Fuchinobe, Sagamihara, 229-8558, Japan}
\affiliation{$^3$ RIKEN Center for Emergent Matter Science (CEMS), Wako, Saitama 351-0198, Japan}

\begin{abstract}
Magnetic skyrmions, vortex-like swirling spin structures with quantized topological number, observed in chiral magnets, are appealing for potential applications in spintronics because it is possible to control their motion with ultralow current density. 
In order to realise skyrmion-based spintronic devices, it is essential to understand the skyrmion motions in confined geometries. 
Here we show by micromagnetic simulations that the current-induced motion of skyrmions in the presence of geometrical boundaries is very different from that in an infinite plane. 
In a channel with finite width, the transverse confinement results in steady state characteristics of the skyrmion velocity as a function of current similar to those of domain walls in ferromagnets, 
whereas the transient behaviour depends on the initial distance of the skyrmion from the boundary. 
In addition, we show that single skyrmions can be created by electric current in a simple constricted geometry of a plate-shaped specimen of suitable size and geometry. 
These findings could guide the design of skyrmion-based devices, in which skyrmions are used as information carriers.
\end{abstract}

\maketitle
\newpage
Skyrmions~\cite{Skyrme62} and their crystallization into a triangular lattice 
have been recently discovered in some ferromagnets without inversion symmetry such 
as MnSi, Fe$_{1-x}$Co$_x$Si, Cu$_2$OSeO$_3$~\cite{Muhlbauer09,Munzer10,Seki12} as 
a consequence of the competition between ferromagnetic exchange coupling $J$ and 
Dzyaloshinskii-Moriya (DM) interaction $D$ under an external magnetic field $\bm{B}$.
In chiral magnets, the ground state for $\bm{B}=0$ is the helical state, which is a successive alignment of Bloch domain walls.
The increase of $B$ changes the helical phase to the skyrmion-crystal phase at $B=B_{\mathrm{c}1}$, and eventually to the ferromagnetic phase at $B=B_{\mathrm{c}2}$.
The skyrmion has magnetization antiparallel (parallel) 
to $\bm{B}$ at their centre (periphery). 
Skyrmion-crystals are experimentally observed both in 
bulk samples~\cite{Muhlbauer09,Munzer10} and 
thin-plate specimens~\cite{YuXZ10,YuXZ11}. 

The energy for each interaction per spin is given by $J/\xi^2$, $D/\xi$, and $B$ for the ferromagnetic exchange interaction, DM interaction, and the Zeeman energy, respectively, 
where $\xi$ is the length scale of the spatial variation of the magnetization. 
The competition among these three interactions determines the phase diagram and size $\xi$ of the spin texture. 
The balance between the former two gives the $\xi_1 \simeq Ja/D$ ($a$: the lattice constant set to be 1 in the followings) and the energy scale $E_1 \simeq D^2/J = (D/J)^2 J \sim (10^{-2}-10^{-4})J$. 
Both $B_{\mathrm{c}1}$ and $B_{\mathrm{c}2}$ are of the order of $E_1$, and the helix period for $B<B_{\mathrm{c}1}$ and the size of the skyrmion for $B_{\mathrm{c}1}<B<B_{\mathrm{c}2}$ are both of the order of $\xi_1$. 
When $B$ is much larger than $E_1$, on the other hand, the balance between the second and third terms determines the length scale as $\xi_2 \simeq D/B$. 
The length scales $\xi_1$ and $\xi_2$ are of the order of 3-100 nm, and are much smaller than the size of magnetic bubbles with a micrometer scale in thin films of ferromagnets induced by the dipolar interaction. 
Note also that the dynamics of magnetic bubbles under current should be identical to that of skyrmions, 
although the main means to manipulate the magnetic bubbles was the magnetic field~\cite{Malozemoff79}.
Moreover, the skyrmion is stable even near the room temperature in certain DM magnets~\cite{YuXZ11}, 
and can be manipulated by much lower electric current than ferromagnetic domain 
walls~\cite{Jonietz10,YuXZ12,Iwasaki13}. 
These properties, that is, nanometric size, high operational temperature, and low threshold current, are advantageous for technical application to high-density storage devices.

In order to use skyrmions as information carriers, we must be able to drive their motion along nanostructures, as well as to nucleate them and annihilate them at will. 
However, the knowledge about motions in constricted geometries is lacking, 
and the generation of skyrmions is recognized to be very difficult because of the topological stability, 
that is, it can never be created or annihilated by continuous variation of 
spin configuration from uniform ferromagnetic state. 
This means that discontinuous flip of local magnetization is necessary for its creation, 
which inevitably results in an energy cost of the order of $J$ 
although a typical energy scale of skyrmion per spin is $E_1 \sim (10^{-4}-10^{-2}) J$ defined above. 
Recently, the method to create skyrmion using a circulating current was proposed~\cite{Han12}. 
In addition, it was reported that creation of skyrmion can be achieved by ultrashort single optical 
laser pulse~\cite{Finazzi13}.
Nucleation of skyrmions from stripe has been also observed in hexaferrite by 
Lorentz microscope~\cite{YuXZ12PNAS}.

In this paper, we study the current-induced dynamics of skyrmion in several kinds of constricted 
geometries by the micromagnetic simulation. 
We have found that the confinement and boundary effect drastically changes the current-induced skyrmion dynamics including the steady-state current-velocity relation, transient phenomena, and also the creation and annihilation of skyrmions.

\section*{Model and simulation}
The spin system of chiral magnets is described by a classical Heisenberg 
model on the 2D square lattice, where the dimensionless local magnetic moments 
$\bm{M}_{\bm{r}}$ defined as 
$\bm{M}_{\bm{r}} \equiv -\bm{S}_{\bm r}/\hbar$ ($\bm{S}_{\bm r}$ is the local spin at 
${\bm r}$ and $\hbar$ is a Plank constant divided by $2\pi$) are treated as classical vectors whose length is fixed to be $|\bm{M}_{\bm r}|=M$.
We consider the impurity effect by introducing magnetic anisotropy at randomly distributed 
impurity sites.
The Hamiltonian is given by
\begin{align}
\mathcal{H}=&-J\sum_{\bm r} {\bm M}_{\bm r} \cdot 
\left( {\bm M}_{{\bm r}+{\bm e_x}}+{\bm M}_{{\bm r}+{\bm e_y}} \right) 
\notag \\
&-D\sum_{\bm r} \left( 
 {\bm M}_{\bm r} \times {\bm M}_{{\bm r}+{\bm e_x}} \cdot {\bm e_x}
+{\bm M}_{\bm r} \times {\bm M}_{{\bm r}+{\bm e_y}} \cdot {\bm e_y} \right) 
\notag \\
&-{\bm B} \cdot \sum_{\bm r} {\bm M}_{\bm r}
-A \sum_{{\bm r}\in I} M_{z{\bm r}}^2.
\label{eqn:Hamiltonian}
\end{align}
Here $\bm{e_x}$ and $\bm{e_y}$ are unit vectors to $x$ and $y$ directions respectively, and $I$ denotes a set of positions of the impurities.
The magnetic field $\bm{B}$=$(0, 0, B)$ is applied normal to the plane, and we adopt $J=1$ meV.
The typical DM interaction $D=0.18J$, which is used in most of this paper, gives the transition between 
the helical and skyrmion-crystal phases at $B_{\mathrm{c}1}=0.0075J$, 
while that between the skyrmion-crystal and ferromagnetic phases at $B_{\mathrm{c}2}=0.0252J$.
These values are appropriate for MnSi as explained in Ref.~\cite{Iwasaki13}.
Anisotropy $A$ at the impurity sites is easy-axis along $M_z$-direction when $A>0$ while it is easy-plane when $A<0$. 

We study the current-induced spin dynamics at $T=0$ by numerically solving 
the Landau-Lifshitz-Gilbert (LLG) equation (see Methods):
\begin{align}
\frac{{\rm d} {\bm M}_{\bm r}}{{\rm d} t}=& 
-\gamma {\bm M}_{\bm r} \times {\bm B}^{\rm eff}_{\bm r}
+\frac{\alpha}{M} {\bm M}_{\bm r} \times 
\frac{{\rm d}{\bm M}_{\bm r}}{{\rm d} t}
+\frac{pa^3}{2eM} \left( {\bm j}(\bm{r}) \cdot {\bm \nabla} \right) {\bm M}_{\bm r} 
\notag \\
&-\frac{pa^3\beta}{2eM^2} \left[ {\bm M}_{\bm r} \times \left( {\bm j}(\bm{r}) \cdot {\bm \nabla} \right) 
{\bm M}_{\bm r} \right],
\label{eqn:LLG}
\end{align}
with ${\bm B}^{\rm eff}_{\bm r}=-\frac{1}{\hbar\gamma}\frac{\partial \mathcal{H}}{\partial {\bm M}_{\bm r}}$.
Here $\gamma$ is the gyromagnetic ratio, $p$ is the spin polarization of 
the electric current, and $e(>0)$ is the elementary charge.
The second term denotes the Gilbert damping where $\alpha$ is fixed at $\alpha=0.04$.
The third and fourth terms describe the coupling between spins and spin-polarized electric 
current $\bm{j}(\bm{r})$; the third (fourth) term describes the coupling via the spin transfer torque
(nonadiabatic effects).
The coefficient of the fourth term $\beta$ determines the strength of the nonadiabatic torque.

The electric current density ${\bm j}({\bm r})$ 
is assumed to be proportional to the electric field 
${\bm E}({\bm r}) = - \nabla \phi({\bm r})$
with $\phi({\bm r})$ being the scalar potential, that is, 
${\bm j}({\bm r})=\nabla \phi({\bm r})$.
Here the coefficient is absorbed into the definition 
of $\phi(\bm{r})$.
For a steady current distribution, the conservation of 
current, $\nabla \cdot \bm{j} =0$, leads to the Poisson equation
for $\phi(\bm{r})$,
$\Delta \phi(\bm{r})=0$, 
which should reflect the system shape 
with boundary conditions that the current density $j$ is flowing in from right and flowing out to left as
\begin{align}
\frac{\partial \phi}{\partial n}= 
\begin{cases}
j & \mbox{(at the left end of sample)} \\
-j & \quad \mbox{(at the right end of sample)} \\
0 & \quad \mbox{(otherwise)}
\end{cases}
\label{eqn:bc}
\end{align}
(see Methods for numerical method to calculate $\bm{j}(\bm{r})$).

\section*{Skyrmion motion in a finite width channel}
\begin{figure}
\includegraphics[scale=1.0]{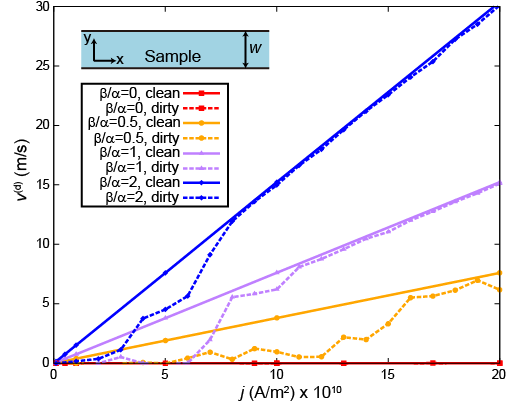}
\caption{}
\label{fig:jv}
\end{figure}

The motion of a skyrmion confined in a narrow region is completely different 
from that in the non-confined plane because of confining forces from boundaries. 
The sample shape considered here is a long stripline-shaped system along $x$-direction 
with nanometric width $w$ along $y$-direction as shown in the inset of Fig.~\ref{fig:jv}.
We set $B=0.0278J$.
The relation between the current density $j$ and the drift velocity $v^{(\mathrm{d})}$ of 
skyrmion in the steady state after the transient time is shown in Fig.~\ref{fig:jv}.
Here the impurity concentration $n$ is fixed at $n=0$ for the clean case, 
while at $n=0.1$ \% for the dirty case.
The impurity strength $A$ is $A=0.2J$ (easy-axis anisotropy).
For the dirty case, we take the average over eight different impurity distributions.
Compared with the universal $j$-$v^{(\mathrm{d})}$ relation of skyrmions without boundary effect~\cite{Iwasaki13}, 
the $j$-$v^{(\mathrm{d})}$ relation in Fig.~\ref{fig:jv} depends strongly on 
$\alpha$, $\beta$ and the impurity effect, which is very similar to that of the helical phase or 
ferromagnetic domain walls.
Especially when $\beta=0$, the skyrmion stops even without impurity, which is regarded 
as the intrinsic pinning (see Supplementary Movie 1).
This characteristic can be explained by the Thiele equation for the centre of mass 
$\bm{R}=(X,Y)$~\cite{Everschor12,Schulz12} of a spin texture
(see Supplementary Information I for details):
\begin{eqnarray}
\bm{G} \times (\bm{v}^{(\mathrm{s})} - \bm{v}^{(\mathrm{d})})
+\bm{\mathcal{D}}(\beta \bm{v}^{(\mathrm{s})} - \alpha \bm{v}^{(\mathrm{d})})
-\nabla V(\bm{r})=\bm{0},
\label{eqn:Thiele}
\end{eqnarray}
where $\bm{v}^{(\mathrm{d})}=\dot{\bm{R}}$ is the drift velocity of the spin texture and 
$\bm{v}^{(\mathrm{s})}=-\frac{p a^3}{2eM}\bm j$ is the velocity of the conduction electrons,
which can be identified as the current density $\bm j$ except a factor.
The first term in the left hand side of Eq.~\eqref{eqn:Thiele} describes the Magnus force
with $\bm{G}=\mathcal{G} {\bm e}_z$ with $\bm{e}_z$ being the unit vector along $z$-direction 
and $\mathcal{G}= 4 \pi Q $ ($Q=-1$: skyrmion number) for the skyrmion,
while $\bm{G}= \bm{0}$ for the domain wall or helix. 
The second term represents the dissipative force where 
the components $\mathcal{D}_{ij}$ of the tensor $\bm{\mathcal{D}}$
are $\mathcal{D}_{xx}= \mathcal{D}_{yy}= \mathcal{D}$ and $0$ for otherwise. 
The third term $-\nabla V(\bm{r})$ stands for the force due to the potential $V(\bm{r})$ from surrounding environment such 
as boundaries as well as that from the impurities.
The mechanism for the repulsive potential due to the boundary is that the in-plane 
tilt at the edge has the opposite direction to that at the perimeter of the skyrmion. 
We have demonstrated that this mechanism produces the potential barrier in the presence of the DM interaction using a one-dimensional model in Supplementary Information I\hspace{-.1em}I.

Let us start with the case of skyrmion motion without the boundary or impurities, i.e., $V=0$.
In this case, considering the fact that $\alpha, \beta \ll 1$  and $\mathcal{D} \sim 1$, 
the dominant term in Eq.~\eqref{eqn:Thiele} is $\bm{G} \times (\bm{v}^{(\mathrm{s})} - \bm{v}^{(\mathrm{d})})$
and hence we obtain $\bm{v}^{(\mathrm{d})} = \bm{v}^{(\mathrm{s})}$, i.e., universal
current-velocity relation.
The small correction due to $\alpha, \beta$ gives the skyrmion Hall effect transverse to the current
given by the additional velocity $\delta \bm{v}^{(\mathrm{d})} = 
\bm{e}_z \times \mathcal{D} (\alpha - \beta) \bm{v}^{(\mathrm{s})}/ \mathcal{G}$.
When the pinning force $\bm{F}_{\rm pin} = - \nabla V_{\rm imp.}$ is there, 
the threshold current density $v^{(\mathrm{s})}_\mathrm{c}$ is determined as
$v^{(\mathrm{s})}_\mathrm{c} \simeq |\bm{F}_{\rm pin} |/|\mathcal{G}|$, which is much smaller than 
the case of domain wall motion where $\bm{G}= \bm{0}$ and hence 
$v^{(\mathrm{s})}_\mathrm{c} \simeq |\bm{F}_{\rm pin} |/(\beta \mathcal{D})$.
 
Now we turn to the confined case. 
In the steady state, the drift velocity $\bm{v}^{(\mathrm{d})}$ is along the $x$-direction
while its $y$-coordinate is nonzero due to the presence of the boundary.
In the absence of impurities, $V$ comes from the confining potential
and hence $\partial_y V$ is finite near the boundary while $\partial_x V=0$.
Therefore, the $x$-component of Eq.~\eqref{eqn:Thiele} reads 
\begin{align}
{v}_x^{(\mathrm{d})}=\frac{\beta}{\alpha}{v}_x^{(\mathrm{s})},
\label{eqn:vel}
\end{align}
while that of $y$-component
\begin{align} 
\mathcal{G} ({v}_x^{(\mathrm{s})} - {v}_x^{(\mathrm{d})}) - \partial_y V =0
\label{eqn:ycomp}
\end{align}
gives the equilibrium condition and determines $Y$.
Namely, the Magnus force is balanced with the confining force,
while the dissipative terms determine the velocity.  
Equation \eqref{eqn:vel} is exactly the same as that of the
domain wall motion. 
In the presence of impurities, the threshold current density 
$v_\mathrm{c}^{(\mathrm{s})}$ is determined by 
as $v^{(\mathrm{s})}_\mathrm{c} \simeq |{F}_{x \rm pin} |/(\beta \mathcal{D})$ , 
which is again very similar to the case of domain wall~\cite{Iwasaki13}. 
Assuming the confining potential as $V=Y^2/(2M)$, Eq.~\eqref{eqn:ycomp} means that 
$Y$ is the ``momentum" conjugate to $X$ with $M$ playing the role of mass.

This might appear to be a serious obstacle for applications. 
However, the distance $l$ a skyrmion travels before it stops can be
long.  (Fig.~\ref{fig:transient} and also see Supplementary Movies 2 and 3).
If the skyrmion starts from the middle of the channel, $l$ 
reaches more than $0.6$ $\mu$m for $\beta/\alpha=0.5$, 
$j=5.0 \times 10^{10}$ A/m$^2$ ($<j_\mathrm{c}^{(\mathrm{pin})}$) and $w=75$ nm 
(this value of $w$ is approximately four times larger than the diameter of skyrmion $\xi \simeq 18$ nm, where the radius is defines as twice the distance from the core $M_z=-1$ to the perimeter $M_z=0$).
The distance $l$ could depend on various factor, for example, type or strength of disorder, 
size of skyrmion, $\alpha, \beta$, and initial position of the skyrmion.
We have investigated other parameter sets and we can conclude that only its initial $y$-coordinate and $\alpha - \beta$ affect the traveling distance $l$ (see Supplementary Information I\hspace{-.1em}I\hspace{-.1em}I).

\section*{Skyrmion nucleation}
\begin{figure}
\includegraphics[scale=1.0]{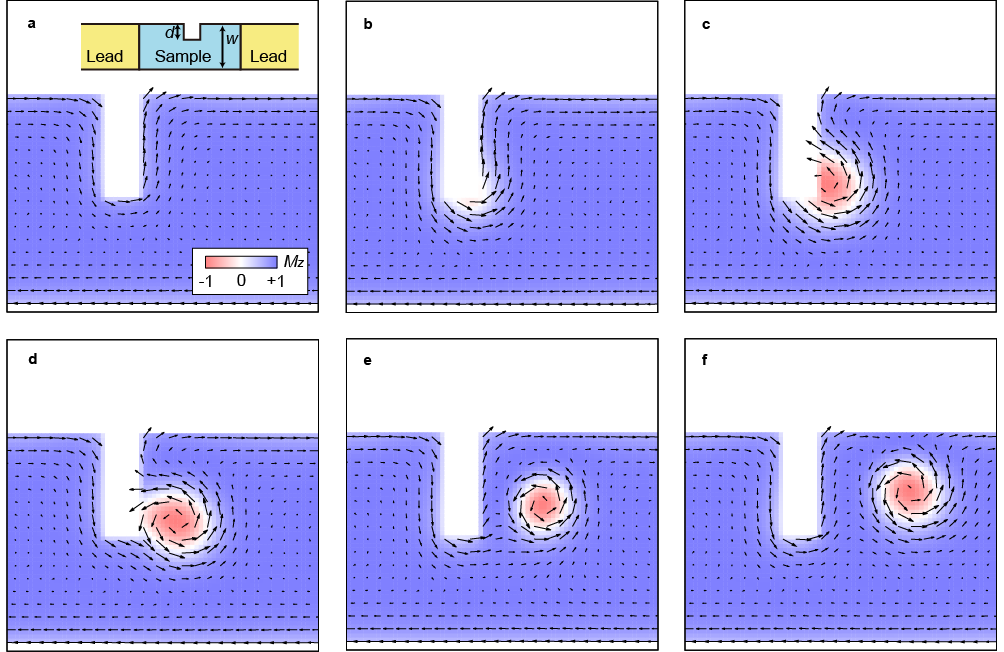}
\caption{}
\label{fig:creation}
\end{figure}

Next we demonstrate the creation of skyrmion in a stripline-shaped system 
with a square notch structure (the inset of Fig.~\ref{fig:creation}).
(The notch is often discussed for the domain wall pinning~\cite{Klaui05,Martinez07,Jang09}, 
and the motion of a skyrmion through a triangular notch has been simulated in~\cite{Fert_rev13})
Figure~\ref{fig:creation} exhibits snapshots of the spin configuration around the notch at selected times, 
where the creation of skyrmion is observed (See also Supplementary Movie 4).
Here we fix the magnetic field at $B=0.0278J$, which is slightly above $B_{\mathrm{c}2}$.
Namely, the ferromagnetic state is the ground state.
Spins without electric current have large 
in-plane components near the boundary owing to the DM interaction 
(Fig.~\ref{fig:creation}a and Supplementary Information I\hspace{-.1em}V).
As the electric current flows, the spin texture at the notch swells out due to the spin transfer torque, 
and one can regard this portion as the seed of a skyrmion (Fig.~\ref{fig:creation}b).
Subsequently spins behind the seed become automatically twisted and pointing down 
due to the DM interaction, and eventually 
the skyrmion core is created (Fig.~\ref{fig:creation}c).
This spin twist toward generating the skyrmion core is due to the spin precession.
The unique direction of the precession breaks the reflection symmetry, and, thereby, an asymmetry 
with respect to the sign of $j$ arises as shown below.
At the early stage after its creation, the radius of the created skyrmion slightly oscillates, 
and eventually the radius converges to arrive at a metastable energy minimum via 
the Gilbert-damping process.

The magnetic structure at the notch, not the winding current pattern, is essential in creation process.
To confirm this, we simulated the dynamics in the following two fictitious conditions.
The first condition is the uniform current distribution without modifying the magnetic structure at the notch.
In this case, skyrmions are created as in the case of realistic current distribution.
The next condition is the modified magnetic structure such that spins next to the edge point in 
$z$-direction by a huge magnetic field.
In this case, skyrmion was never created. 

\begin{figure}
\includegraphics[scale=1.0]{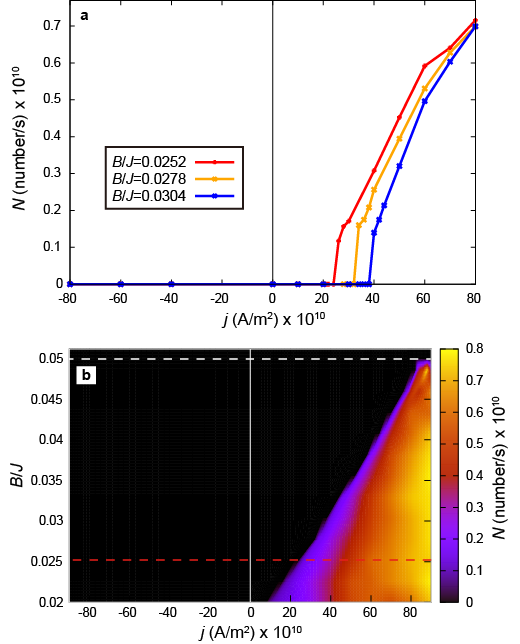}
\caption{}
\label{fig:cre_rate}
\end{figure}

The creation rate of skyrmions $N$ (the number of skyrmions created per time) as a function of the 
current density $j$ for several values of magnetic field $B$ is displayed in Fig.~\ref{fig:cre_rate}a.
In Fig.~\ref{fig:cre_rate}b, a colour plot of the rate $N$ is presented in the plane of $j$ and $B$.
We find that a larger current density is necessary to create skyrmions under a stronger magnetic field.
Surprisingly, skyrmions are created even under a magnetic field which 
largely exceeds $B_{\mathrm{c}2}$ up to $B_\mathrm{c}$, 
where the skyrmion ceases to be metastable.
In addition, the $j$-$N$ plot is asymmetric between $j$ and $-j$, i.e.,  
the current flowing in the opposite direction cannot generate skyrmions.
The threshold current density $j_\mathrm{c}^{(\mathrm{cr})}$ is of the order of 
$\sim10^{11}$--$10^{12}$ A/m$^2$. 
The energy cost to generate a skyrmion is 
$\left| \frac{\partial \mathcal{H}}{\partial \bm{M}} \right| \sim J$, 
and this energy must be supplied by the electric current.
Therefore, the spin transfer torque in the LLG equation should be of the same order as 
the precession term, namely, 
$\left| -\gamma \bm{M} \times \bm{B}^\mathrm{eff} \right|
\sim \left| \frac{pa^3}{2eM} \left( \bm{j} \cdot {\bm \nabla} \right) {\bm M} \right|$, 
which leads to
\begin{align}
j &\sim \frac{eJ}{\hbar a^2} \simeq 10^{12} \ \mbox{A/m$^2$},
\end{align}
which is consistent with the obtained $j_\mathrm{c}^{(\mathrm{cr})}$.

The creation rate $N$ depends on various conditions.
If the depth of the notch $d$ is too small (relative to the skyrmion radius), 
only a portion of a skyrmion can appear inside the sample, so that 
the swirling of spins to constitute a skyrmion structure cannot be completed, 
and eventually it disappears as time goes by.
On the other hand, if $d$ is so large that $w-d$ is small ($w$: the sample width), the creation 
never occurs.
The angle of the notch corner $\theta$ is another issue.
We examined five angles, $\theta=30^\circ,45^\circ,90^\circ,120^\circ$ and $135^\circ$, 
and found that $90^\circ$ is the most suitable for the creation.
We have also studied the case of rounded notch, and find that the
skyrmion creation occurs when the curvature radius is comparable to the
size of the skyrmion. This means that the sharp edge at the 
corner of the notch is not essential, while the in-plane components 
of the spins along the notch 
and their shift due to the spin transfer torque are the 
essential mechanism. 
It is noted that a sign change of the DM interaction does not alter 
the $N$-$j$ plot (Fig.~\ref{fig:cre_rate}a), 
while that of $B$ exchanges $j$ and $-j$ in Figs.~\ref{fig:cre_rate}a and b.

\section*{Skyrmion dynamics at the edge of magnetic material}
\begin{figure}
\includegraphics[scale=1.0]{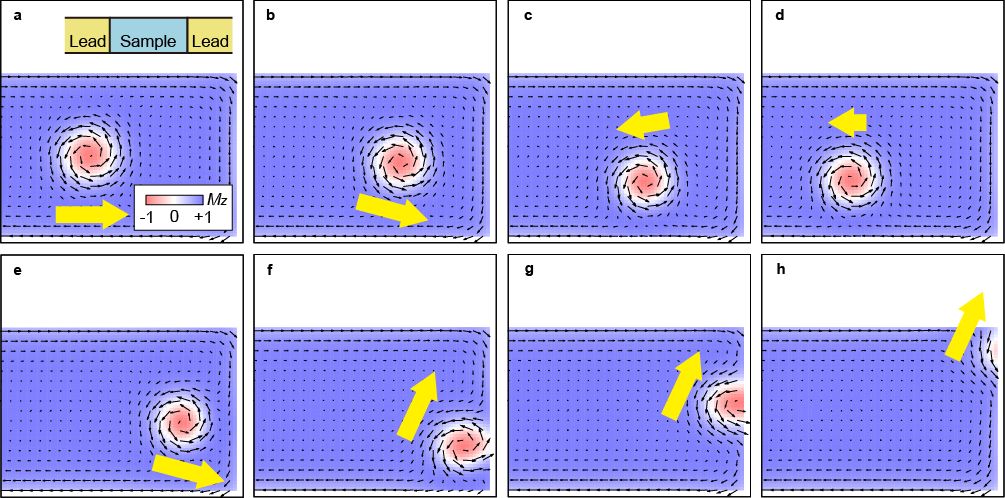}
\caption{}
\label{fig:annihilation}
\end{figure}

Finally we study the motion at a junction of magnetic region and nonmagnetic leads.
The sample considered here is shown in the inset of Fig.~\ref{fig:annihilation}. 
We set $B=0.0278J$ and study only the clean case ($n=0$).
Two types of skyrmion dynamics near the boundary are presented in Fig.~\ref{fig:annihilation}.
The case with a small current density is shown in Figs.~\ref{fig:annihilation}a--d 
(See also Supplementary Movie 5).
The skyrmion bounces and cannot reach the boundary since it cannot overcome 
a repulsive potential from the boundary.
In this bouncing process, the repulsive potential induces a motion 
transverse to the boundary because of the Magnus force.
Eventually the skyrmion stops at a position slightly below the central line of the system.
The larger current density enables a skyrmion to overcome the potential barrier, and pushes 
the skyrmion to the sample edge, resulting in annihilation of the skyrmion 
(Figs.~\ref{fig:annihilation}e--h, and also Supplementary Movie 6).
Since the skyrmion is subject to repulsive and attractive potentials before and after it overcomes 
the potential barrier, respectively, its trajectory is curved first downward and then upward.
The threshold current density $j_\mathrm{c}^{(\mathrm{T})}$ between the above two motions is 
$j_\mathrm{c}^{(\mathrm{T})} \simeq 1.5 \times 10^{11}$ A/m$^2$ for $\beta/\alpha=1$.
However, the strength of boundary barrier can depend on the system parameters.
The additional calculations revealed that the size of the skyrmion, which is controlled 
by the ratio $D/B$, is the key factor to determine the threshold current density,
and smaller skyrmion can overcome the barrier more easily (see Supplementary Information I\hspace{-.1em}I\hspace{-.1em}I).
The detailed analysis based on Eq.~\eqref{eqn:Thiele} is given in Supplementary Information I.

\section*{Discussion}
Some remarks are in order on the magnetic vortex, which has the similar dynamics to 
skyrmion and intensively studied~\cite{Shibata06,Guslienko06,Liu07,Wysin96,Ono07}.
The vector ${\bm G}=\mathcal{G} {\bm e}_z$ in eq.(\ref{eqn:Thiele}) for a magnetic vortex is 
given by $\mathcal{G}= 2 \pi p q$ where $p= \pm1$ specifies the direction of the spin at centre,
while $q= \pm1$ is the vorticity. Although the equation of motion looks similar to skyrmion case,
there are several essential differences. 
(i) The vortex is a non-local object since the spins  far away from the
centre are winding within the $xy$-plane, which causes the logarithmic divergence of the energy with 
respect to the size of the sample~\cite{Wysin96}. 
Therefore, the current-driven motion of a vortex is often studied in 
a finite size sample of e.g. disk shape. This is in sharp contrast to skyrmion, which has the 
finite size and regarded as an independent ``particle". 
(ii) For the magnetic vortex realised in a disk-shaped sample due to the dipolar interaction, 
there are four degenerate states corresponding to $p= \pm 1,q= \pm 1$. 
It has been demonstrated that $p$ can change during the current-driven motion, 
which results in the reversal of the direction of rotational motion~\cite{Ono07}.
In the case of skyrmion studied in this paper, the stable skyrmion state is unique 
with the skyrmion number $Q$ being determined by the direction of the external magnetic field, 
and hence its dynamics is stable.

In conclusion, we have found that the $j$-$v^{(\mathrm{d})}$ relation of a confined skyrmion is similar 
to that of ferromagnetic domain walls, and have 
demonstrated a new way to create and annihilate skyrmions in constricted geometries.
The creation and annihilation can be controlled easily by an electric current and a magnetic field.

\section*{Methods}
{\bf Numerical Simulation of the LLG equation.}
We use the fourth-order Runge-Kutta method to solve the LLG equation.
To study the $j$-$v^{(\mathrm{d})}$ relation and the skyrmion creation, 
the periodic boundary condition is imposed at the left 
and right sides of the sample, while the open boundary condition at other boundaries.
To study the skyrmion annihilation, the open boundary condition is imposed at all boundaries. 
The natural units of time $t$ and current density $j$ are $\tau \equiv \hbar/J$ and 
$\kappa \equiv \frac{2eMJ}{pa^2\hbar}$, respectively.
With a typical lattice constant $a=5$ {\AA}, spin-polarization $p=0.2$ and the magnitude of local magnetic 
moment $M=1$, the values of $\tau$ and $\kappa$ become $\tau \simeq 6.5 \times 10^{-13}$ s and 
$\kappa \simeq 1.0 \times 10^{13}$ A/m$^2$.
We use these values to convert the units of simulated current density and time.

{\bf Calculation of current distribution}
To solve the Poisson equation $\Delta \phi(\bm{r})=0$ with the Neumann-type boundary 
conditions \eqref{eqn:bc}, we employ the finite element method~\cite{FEM}.
The system is divided up into triangular meshes by the following way.
We first draw lines between neighboring sites.
In this step, the system is divided into squares of the same size.
Next we draw a line from the left bottom point to the right top point for each square 
to complete the triangulation.

\section*{Acknowledgements}
The authors are grateful for insightful discussions with Y. Tokura, M. Kawasaki and X. Z. Yu. 
This work was supported by Grant-in-Aids for Scientific Research (Nos.~24224009, 25870169, 25287088) from the Ministry of Education, Culture, Sports, Science and Technology (MEXT) of Japan, 
Strategic International Cooperative Program (Joint Research Type) from Japan Science and Technology Agency, 
and by Funding Program for World-Leading Innovative R\&D on Science and Technology (FIRST Program). 
M. M. was supported by the G-COE Program ``Physical Sciences Frontier" from MEXT of Japan.

\section*{Author Contributions}
J. I. performed the numerical calculations. J. I., M. M., and N. N. contribute in analyzing the data and writing the paper.

\section*{Competing Financial Interests}
The authors declare that they have no competing financial interests.

\section*{Correspondence}
Correspondence and requests should be addressed to J. I. (iwasaki@appi.t.u-tokyo.ac.jp) and N. N. (nagaosa@riken.jp).

\section*{Figure Legends}
\begin{enumerate}
\item {\bf Steady state velocity of the current-induced motion of skyrmion after the transient time in a finite width channel:}
Drift velocity $v^{(\mathrm{d})}$ of the current-induced steady-state motion of skyrmion 
in a finite system as a function of the current density $j$ for several values of $\beta/\alpha$, 
where $\alpha$ and $\beta$ are respectively the coefficients of the Gilbert damping and the nonadiabatic effect.
For the dirty case with impurity concentration of $n=0.1$ \%, 
the values averaged over eight different patterns of impurity distributions are presented.
Note that the $j$-$v^{(\mathrm{d})}$ characteristic for $\beta/\alpha=1$ and $n=0$ is identical to the universal relation for any spin texture at $\alpha = \beta$, which is realised for skyrmion-crystal in a non-confined space even with $\alpha \ne \beta$~\cite{Iwasaki13}. 
The inset is a schematic picture of a channel of the width $w$.

\item {\bf Transient behavior of the current-induced motion of skyrmion in a wide channel:}
Snapshots of skyrmion motion in the transient period before it stops beside the boundary in a stripline-shaped system with the impurity concentration of $n=0.1$ \%.
The width of the channel $w$ is $150$ sites while the diameter of the skyrmion is $\sim 35$ sites.
The radius of skyrmion is defined as twice the distance from the core $M_z=-1$ (red colour) to the perimeter $M_z=0$ (white colour).
Colour plot represents the $z$-components of the magnetic moments.
The numerical simulation is performed for $\beta/\alpha=0.5$ 
with current density $j=5.0 \times 10^{10}$ A/m$^2$, which is below the threshold value in this constricted geometry while it is well above the critical value in the infinite space.
Positions of the impurities are indicated by green dots.
Times corresponding to respective skyrmion positions are
{\bf a,} $t=0$, 
{\bf b,} $t=2.08 \times 10^{-8}$ s, 
{\bf c,} $t=4.16 \times 10^{-8}$ s, 
{\bf d,} $t=6.24 \times 10^{-8}$ s, 
{\bf e,} $t=8.32 \times 10^{-8}$ s, 
{\bf f,} $t=1.04 \times 10^{-7}$ s, 
{\bf g,} $t=1.25 \times 10^{-7}$ s, 
{\bf h,} $t=1.46 \times 10^{-7}$ s, 
and {\bf i,} $t=1.77 \times 10^{-7}$ s.
The skyrmion are pinned after the configuration {\bf i}.

\item {\bf Creation process of skyrmion:}
Snapshots of dynamical spin configurations at selected times in creating a skyrmion for $j=3.6 \times 10^{11}$ A/m$^2$.
In-plane components of the magnetic moments at sites $(i_x,i_y)$ are represented by arrows when 
$\bmod(i_x, 3)=\bmod(i_y, 3)=1$.
Colour plot represents the $z$-components of the magnetic moments.
Times corresponding to respective figures are 
{\bf a,} $t=0$, 
{\bf b,} $t=9.10 \times 10^{-11}$ s, 
{\bf c,} $t=18.85 \times 10^{-11}$ s, 
{\bf d,} $t=28.60 \times 10^{-11}$ s, 
{\bf e,} $t=38.35 \times 10^{-11}$ s, 
and {\bf f,} $t=48.10 \times 10^{-11}$ s.
The inset is a schematic picture of the system with a notch structure for skyrmion creation.

\item {\bf Rates of skyrmion creation:}
{\bf a,} Creation rate $N$ of skyrmion as a function of the current density $j$ 
for several values of magnetic field $B$.
{\bf b,} Colour plot of the creation rate $N$ in the plane of current density $j$ and magnetic field $B$. 
The white dashed line indicates the critical magnetic field $B_\mathrm{c}/J$, above which the system cannot hold a metastable skyrmion state.
The red dashed line indicates the critical magnetic field $B_{\mathrm{c}2}/J$ 
for the phase transition between skyrmion-crystal and ferromagnetic phases in the ground state. 

\item {\bf Two types of skyrmion motions near the edge of magnetic region:}
Snapshots of dynamical spin configurations at selected times 
near the end of magnetic region for two different current densities $j$.
In-plane components of the magnetic moments at sites $(i_x,i_y)$ are represented by arrows when 
$\bmod(i_x, 3)=\bmod(i_y, 3)=1$.
Colour plot represents the $z$-components of the magnetic moments.
The yellow arrows indicate the direction and the magnitude of velocity at each moment.
The current density $j$ is $j=1.0 \times 10^{11}$ for {\bf a}--{\bf d} 
and $j=3.0 \times 10^{11}$ for {\bf e}--{\bf h}.
Times corresponding to respective figures are 
{\bf a,} $t=1.95 \times 10^{-9}$ s,
{\bf b,} $t=2.28 \times 10^{-9}$ s,
{\bf c,} $t=2.60 \times 10^{-9}$ s,
{\bf d,} $t=2.93 \times 10^{-9}$ s,
{\bf e,} $t=8.13 \times 10^{-10}$ s,
{\bf f,} $t=9.10 \times 10^{-10}$ s,
{\bf g,} $t=9.75 \times 10^{-10}$ s,
and {\bf h,} $t=10.08 \times 10^{-10}$ s.
The inset is a schematic picture of the system with junctions of magnetic region and nonmagnetic leads.
\end{enumerate}

\newpage
\section*{Supplementary Information}
is linked to the on-line version of the paper at www.nature.com/nature/

\begin{enumerate}
\item {\bf Supplementary Movie 1:}
Intrinsic pinning of a skyrmion motion in a stripline-shaped system without impurity.
Final $y$-coordinate of the skyrmion is shifted from the central line of the channel.
The width of the system $w$ is $50$ sites while the diameter of the skyrmion is $\sim 35$ sites.
In-plane components of the magnetic moments at sites $(i_x,i_y)$ are represented by arrows when 
$\bmod(i_x, 3)=\bmod(i_y, 3)=1$ for a time range from $t=0$ to $t=3.12 \times 10^{-8}$ s.
The numerical simulation is performed for $\beta/\alpha=0$ 
with current density $j=1.0 \times 10^{11}$ A/m$^2$.

\item {\bf Supplementary Movie 2:}
Skyrmion motion in a stripline-shaped system with the impurity concentration of $n=0.1$ \%.
The width of the channel $w$ is $50$ sites while the diameter of the skyrmion is $\sim 35$ sites.
In-plane components of the magnetic moments at sites $(i_x,i_y)$ are represented by arrows when 
$\bmod(i_x, 3)=\bmod(i_y, 3)=1$ for a time range from $t=0$ to $t=8.32 \times 10^{-8}$ s.
The numerical simulation is performed for $\beta/\alpha=0.5$ 
with current density $j=5.0 \times 10^{10}$ A/m$^2$.
Positions of the impurities are indicated by green dots.

\item {\bf Supplementary Movie 3:}
Skyrmion motion in a stripline-shaped system with the impurity concentration of $n=0.1$ \%.
The width of the channel $w$ is $150$ sites while the diameter of the skyrmion is $\sim 35$ sites.
In-plane components of the magnetic moments at sites $(i_x,i_y)$ are represented by arrows when 
$\bmod(i_x, 3)=\bmod(i_y, 3)=1$ for a time range from $t=0$ to $t=2.50 \times 10^{-7}$ s.
The numerical simulation is performed for $\beta/\alpha=0.5$ 
with current density $j=5.0 \times 10^{10}$ A/m$^2$.
Positions of the impurities are indicated by green dots.

\item {\bf Supplementary Movie 4:}
Creation of skyrmion with a designated notch.
In-plane components of the magnetic moments at sites $(i_x,i_y)$ are represented by arrows when 
$\bmod(i_x, 3)=\bmod(i_y, 3)=1$.
Colour plot represents the $z$-components of the magnetic moments.
The numerical simulation is performed with current density $j=3.6 \times 10^{11}$ A/m$^2$ 
for a time range from $t=0$ to $t=6.5 \times 10^{-10}$ s.

\item {\bf Supplementary Movie 5:}
Bouncing of skyrmion at the end of magnetic region.
In-plane components of the magnetic moments at sites $(i_x,i_y)$ are represented by arrows when 
$\bmod(i_x, 3)=\bmod(i_y, 3)=1$.
Colour plot represents the $z$-components of the magnetic moments.
The numerical simulation is performed with current density $j=1.0 \times 10^{11}$ A/m$^2$ 
for a time range from $t=1.63 \times 10^{-9}$ s to $t=3.25 \times 10^{-9}$ s.

\item {\bf Supplementary Movie 6:}
Annihilation of skyrmion at the end of magnetic region.
In-plane components of the magnetic moments at sites $(i_x,i_y)$ are represented by arrows when 
$\bmod(i_x, 3)=\bmod(i_y, 3)=1$.
Colour plot represents the $z$-components of the magnetic moments.
The numerical simulation is performed with current density $j=3.0 \times 10^{11}$ A/m$^2$ 
for a time range from $t=4.8 \times 10^{-10}$ to $t=13.0 \times 10^{-10}$ s.
\end{enumerate}

\newpage

\begin{figure}
\includegraphics[scale=1.0]{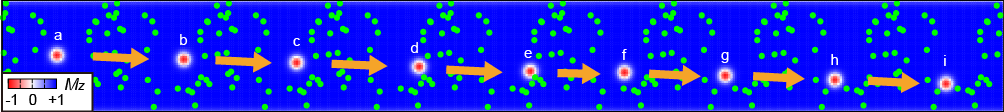}
\caption{}
\label{fig:transient}
\end{figure}

\begin{thebibliography}{99}
\bibitem{Skyrme62} Skyrme, T. H. R. A unified field theory of mesons and baryons. {\it Nuc. Phys.} {\bf 31,} 
556--569 (1962).

\bibitem{Muhlbauer09} M\"uhlbauer, S. et al. Skyrmion lattice in a chiral magnet. {\it Science} 
{\bf 323,} 915--919 (2009).

\bibitem{Munzer10} M\"unzer, W., et al. Skyrmion lattice in the doped semiconductor 
Fe$_{1-x}$Co$_x$Si. {\it Phys. Rev. B} {\bf 81,} 041203(R) (2010).

\bibitem{Seki12} Seki, S., Yu, X. Z., Ishiwata, S. \& Tokura, Y. Observation of skyrmions in a multiferroic 
material. {\it Science} {\bf 336,} 198--201 (2012).

\bibitem{YuXZ10} Yu, X. Z. et al. Real-space observation of a two-dimensional skyrmion crystal. 
{\it Nature} {\bf 465,} 901--904 (2010).

\bibitem{YuXZ11} Yu, X. Z. et al. Near room-temperature formation of a skyrmion crystal 
in thin-films of the helimagnet FeGe. {\it Nature Mater.} {\bf 10,} 106--109 (2011).

\bibitem{Malozemoff79} Malozemoff, A. P. \& Slonczewski, J.C. {\it Magnetic 
Domain Walls in Bubble Materials} (Academic Press, New York, 1979).

\bibitem{Jonietz10} Jonietz, F. et al. Spin transfer torques in MnSi at ultralow current densities. 
{\it Science} {\bf 330,} 1648--1651 (2010).

\bibitem{YuXZ12} Yu, X. Z. et al. Skyrmion flow near room temperature in an 
ultralow current density. {\it Nat. Commun.} {\bf 3,} 988 (2012).

\bibitem{Iwasaki13} Iwasaki, J., Mochizuki, M. \& Nagaosa, N. Universal current-velocity 
relation of skyrmion motion in chiral magnets. {\it Nat. Commun.} {\bf 4,} 1463 (2013).

\bibitem{Han12} Tchoe, Y. \& Han, J. H. Skyrmion generation by current. {\it Phys. Rev. B} {\bf 85,} 174416 (2012)

\bibitem{Finazzi13} Finazzi, M., et al. Laser-induced magnetic nanostructures with 
tunable topological properties. {\it Phys. Rev. Lett.} {\bf 110,} 177205 (2013).

\bibitem{YuXZ12PNAS} Yu, X. Z. et al. Magnetic stripes and skyrmions with helicity reversals. 
{\it Proc. Natl. Acad. Sci. USA} {\bf 109,} 8856 (2012).

\bibitem{Everschor12} Everschor, K. et al. Rotating skyrmion lattices by spin torques and field 
or temperature gradients. {\it Phys. Rev. B} {\bf 86,} 054432 (2012).

\bibitem{Schulz12} Schulz, T. et al. Emergent electrodynamics of skyrmions in a chiral magnet. 
{\it Nature Phys.} {\bf 8,} 301--304 (2012).

\bibitem{Klaui05} Kl\"aui, M., et al. Direct observation of domain-wall pinning at nanoscale 
constrictions. {\it Appl. Phys. Lett.} {\bf 87,} 102509 (2005).

\bibitem{Martinez07} Martinez, E., et al. Thermal effects on domain wall depinning 
from a single notch. {\it Phys. Rev. Lett.} {\bf 98,} 267202 (2007).

\bibitem{Jang09} Jang, Y., et al. Current-induced domain wall nucleation and its pinning 
characteristics at a notch in a spin-valve nanowire. {\it Nanotechnology} {\bf 20,} 125401 (2009).

\bibitem{Fert_rev13} Fert, A., Cros, V. \& Sampaio, J. Skyrmions on the track. {\it Nature Nanotech.} {\bf 8,} 152--156 (2013).

\bibitem{Shibata06} Shibata, J., et al. Current-induced magnetic vortex motion 
by spin-transfer torque. {\it Phys. Rev. B} {\bf 73,} 020403 (2006).

\bibitem{Guslienko06} Guslienko, K., et al., Magnetic vortex core dynamics in cylindrical 
ferromagnetic dots. {\it Phys. Rev. Lett.} {\bf 96,} 067205 (2006).

\bibitem{Liu07} Liu, Y., et al. Current-induced magnetic vortex core switching in a 
Permalloy nanodisk. {\it Appl. Phys. Lett.} {\bf 91,} 112501 (2007).

\bibitem{Wysin96} Wysin, G. Magnetic vortex mass in two-dimensional easy-plane magnets. 
{\it Phys. Rev. B} {\bf 54,} 15156--15162.

\bibitem{Ono07} Yamada, K., et al. Electrical switching of the vortex core in a magnetic disk. 
{\it Nature Mater.} {\bf 6,} 269--273 (2007).

\bibitem{FEM} Zienkiewicz, O. C., Taylor, R. L. \& Zhu, J. Z. 
{\it The Finite Element Method : Its Basis and Fundamentals} (Elsevier Butterworth-Heinemann, Oxford, 2005).

\end{thebibliography}
\end{document}